\documentclass[twocolumn,aps,prb,10pt]{revtex4-2}

\usepackage{amsmath}
\usepackage{amssymb}
\usepackage{wasysym}
\usepackage{graphicx}
\usepackage{hyperref}
\usepackage{dsfont}
\usepackage{newtxtext}
\usepackage[varvw]{newtxmath}
\usepackage{mathtools}
\usepackage[shortlabels]{enumitem}

\hypersetup{
    colorlinks,
    linkcolor={blue},
    citecolor={blue},
    urlcolor={blue}
}

\graphicspath{{./}{./figs/}}

\newcommand{\rme}{\mathrm{e}}
\newcommand{\rmi}{\mathrm{i}}

\newcommand{\NCTO}{Na$_2$Co$_2$TeO$_6$}

\newcommand{\Tstar}{{T_{\star}}}
\newcommand{\Tc}{{T_{\mathrm{c}}}}

\DeclareMathOperator{\sgn}{sgn}
\newcommand{\llangle}{\langle\!\langle}
\newcommand{\rrangle}{\rangle\!\rangle}

\begin{document}

\title{%
Ferrimagnetism from triple-q order in Na$_2$Co$_2$TeO$_6$
}

\author{Niccol\`o Francini}
\author{Lukas Janssen}

\affiliation{Institut f\"ur Theoretische Physik and W\"urzburg-Dresden Cluster of Excellence ct.qmat, TU Dresden, 01062 Dresden, Germany}

\begin{abstract}
The candidate Kitaev magnet Na$_2$Co$_2$TeO$_6$ exhibits a characteristic ferrimagnetic response at low temperatures, with a finite residual magnetization that changes sign at a compensation point located at around half the ordering temperature. We argue that the behavior can be naturally understood to arise in this material as a consequence of a noncollinear triple-$\mathbf{q}$ magnetic ground state. Using large-scale classical Monte Carlo simulations, we study the finite-temperature response of the pertinent honeycomb Heisenberg-Kitaev-$\Gamma$-$\Gamma'$ model in weak training fields. Our model features all symmetry-allowed nearest-neighbor exchange interactions, as well as sublattice-dependent next-nearest-neighbor interactions, consistent with the reported crystal structure of the material. We also consider a six-spin ring exchange perturbation, which allows us to tune between the two different magnetic long-range orders that have been suggested for this material in the literature, namely, a collinear single-$\mathbf{q}$ zigzag state and a noncollinear triple-$\mathbf{q}$ state. We demonstrate that the experimentally-observed ferrimagnetic response of Na$_2$Co$_2$TeO$_6$ can be well described within our modeling if the magnetic ground state features noncollinear triple-$\mathbf{q}$ order. The observation of a compensation point, where the residual magnetization reverses sign, suggests a sublattice $g$-factor anisotropy, with a larger out-of-plane $g$-factor on the sublattice with stronger antiferromagnetic intrasublattice exchange. By contrast, a classical Heisenberg-Kitaev-$\Gamma$-$\Gamma'$-type model with collinear zigzag ground state is insufficient even in principle to describe the observed behavior. Our results illustrate the unconventional physics of noncollinear magnetic long-range orders hosted by frustrated magnets with bond-dependent interactions.
\end{abstract}

\date{December 6, 2024}

\maketitle

\section{Introduction}
\label{sec:intro}

Quantum magnets with bond-dependent interactions can host a variety of exotic phases of matter, such as unconventional magnetic long-range orders or long-range-entangled topological orders characterized by fractionalized excitations and emergent gauge fields.
In this context, honeycomb cobaltates with $d^7$ electronic configurations~\cite{liu18, sano18, liu20, winter22, rousochatzakis24} have recently attracted significant attention as possible platforms for the realization of the spin-$1/2$ Kitaev honeycomb model~\cite{kitaev06}.
Candidate materials include \NCTO~\cite{yao20, songvilay20, lin21, chen21, lee21, hong21, samarakoon21, kim22, mukherjee22, sanders22, yang22, yao22, krueger23, yao23, xiang23, zhang23, hong23, pilch23, bera23, gillig23, miao24, zhou24, arneth24, lin24}, Na$_3$Co$_2$SbO$_6$~\cite{yan19, songvilay20, kim22, sanders22, li22, gu24, hu24, miao24}, and BaCo$_2$(AsO$_4$)$_2$~\cite{zhong20, shi21, zhang22, maksimov22b}.
All of the above compounds host additional interactions beyond the nearest-neighbor Kitaev exchange, stabilizing a magnetically-ordered ground state at the lowest temperatures in the absence of an external magnetic field. 
For various cobaltates, the nature of this long-range-ordered state has been a matter of considerable debate.
For instance, neutron scattering experiments on powder samples of \NCTO\ have been interpreted in terms of a collinear single-$\mathbf{q}$ zigzag ground state~\cite{songvilay20, lin21, kim22, sanders22}, while the symmetries observed in single-crystal measurements have been argued to point to a noncollinear triple-$\mathbf{q}$ order~\cite{chen21, krueger23}.
Similarly, electric polarization experiments, designed to distinguish between the two different suggested spin structures, have found ambiguous results: While no measurable electric polarization has been found by one group~\cite{zhang23}, a different group recently reported a finite magnetoelectric response~\cite{kocsis24}.
A similar ambiguity between single- and multi-$\mathbf q$ states appears in Na$_3$Co$_2$SbO$_6$~\cite{kim22, li22, hu24, gu24}.
At finite temperatures and/or in external magnetic fields, the rich phase diagram of cobaltates shows several phase transitions and intermediate magnetic phases whose precise natures also remain still controversial~\cite{lin21, chen21, lee21, hong21, zhong20, zhang23, yao23, xiang23, hong23, bera23, francini24}.

\begin{figure}[b!]
\includegraphics[width=\linewidth]{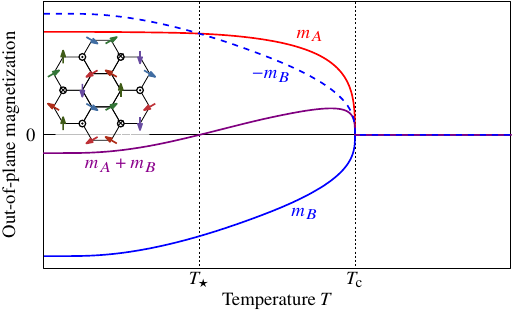}
\caption{Schematic behaviors of out-of-plane sublattice magnetizations $m_A$ (red) and $m_B$ (blue) and net magnetization $m_A + m_B$ (purple) as function of temperature $T$, as found in the present work. The finite magnetization of each crystallographic sublattice is a consequence of the triple-$\mathbf{q}$ ground state, illustrated in the inset. Upon decreasing temperature, one of the sublattice magnetizations ($m_A$) initially grows faster below the critical temperature $\Tc$ than the other one ($m_B$), but saturates at a smaller absolute value in the low-temperature limit. This can be understood to arise from an interplay between different intrasublattice exchange interactions and $g$-factor sublattice anisotropies. The net magnetization $m_A + m_B$ reverses sign at a compensation point $\Tstar$, in agreement with experiments~\cite{yao20}.}
\label{fig:schematic}
\end{figure}

A particularly intriguing phenomenon observed in \NCTO\ is the characteristic ferrimagnetic response observed after cooling the material in weak magnetic training fields in the out-of-plane direction~\cite{yao20}. 
When the training field is removed, a small, but finite, residual magnetization with a \emph{negative} component of the order of $-0.01\mu_\text{B}$/Co$^{2+}$ along the training direction (i.e., opposite to it) remains at the lowest temperatures, as shown schematically in Fig.~\ref{fig:schematic} (purple curve).
Upon increasing the temperature, the magnetization component gradually shifts towards zero and eventually changes sign at a compensation point $\Tstar \simeq 12.5$~K. Above the compensation point, the magnetization component first increases up to a maximum of around $+0.01\mu_\text{B}$/Co$^{2+}$, located only slightly below the antiferromagnetic ordering temperature $\Tc \simeq 26.7$~K, and then drops to zero as $T$ approaches $\Tc$ from below. In the paramagnetic phase for $T > \Tc$, the magnetization vanishes completely.
As has been argued previously~\cite{yao20}, collinear zigzag order features no net moment on either of the sublattices, and as such by itself alone cannot cause ferrimagnetism. Supplementing the zigzag order by weak N\'eel order has been discussed as a possible solution; however, no effective spin model in the relevant extended Heisenberg-Kitaev parameter space, featuring such coexisting order, is known to date~\cite{fouet01, chaloupka10, chaloupka13, chaloupka15, rau14a, rau14b, janssen16, janssen17, consoli20, chern17, stavropoulos23, winter17a, janssen19, rousochatzakis24}.
Furthermore, the refinement of powder neutron diffraction data did not show evidence for such coexisting N\'eel order~\cite{lefrancois16}, which in principle would be revealed by a weak $\mathbf q = 0$ Bragg peak~\footnote{%
A very weak $\mathbf{q} = 0$ Bragg peak below the measurement resolution can of course not be excluded experimentally.}.

In this work, we demonstrate that the characteristic ferrimagnetic response of \NCTO\ can be naturally understood within an effective honeycomb Heisenberg-Kitaev-$\Gamma$-$\Gamma'$ spin model that features a noncollinear triple-$\mathbf{q}$ ground state. Using large-scale classical Monte Carlo simulations, we investigate the finite-temperature response of the model in weak out-of-plane training fields.
In the spin-orbit-coupled system, the SU(2) spin rotation symmetry is reduced to a discrete $C_3^*$ symmetry, which involves $2\pi/3$ pseudospin rotations around the out-of-plane axis combined with $2\pi/3$ lattice rotations~\cite{janssen16, janssen19}.
This allows for four different types of bilinear spin exchange interactions, namely the usual Heisenberg exchange parametrized by $J$, the Kitaev $K$, as well as the off-diagonal $\Gamma$ and $\Gamma'$ interactions~\cite{chaloupka10, chaloupka13, rau14a, rau14b}.
We employ a model that features all four symmetry-allowed interactions on the nearest-neighbor level.
On the next-nearest-neighbor level, we restrict ourselves to Heisenberg interactions, which has previously been shown to be sufficient to reproduce well the low-temperature magnetic excitation spectrum~\cite{krueger23}. Importantly, the reported crystal symmetry of \NCTO~\cite{yao20} allows different next-nearest-neighbor interactions $J_2^A$ and $J_2^B$ on the two different sublattices $A$ and $B$, respectively, which turns out to be crucial in order to understand the ferrimagnetic response of the material.
Furthermore, we consider a six-spin ring exchange interaction, which occurs as leading correction to the nearest-neighbor Heisenberg exchange in the strong-coupling expansion of the honeycomb-lattice Hubbard model~\cite{yang12}, and has previously been successfully employed to model different experimental features of \NCTO~\cite{krueger23, wang23, francini24}.
Here, we use it to easily tune between the collinear single-$\mathbf q$ zigzag order and the noncollinear triple-$\mathbf q$ order in the modeling.

Ferrimagnetism corresponds to a phenomenon in nominally antiferromagnets, in which the magnetic moments on inequivalent sublattices do not fully compensate each other~\cite{kimsekwon22}.
The mechanism that we propose for the ferrimagnetic response of \NCTO\ with a compensation point $\Tstar$~\cite{yao20} is based on three preconditions: 
(1)~different sizes of the magnetic moments, i.e., different $g$-factors, on the two crystallographically inequivalent sublattices, 
(2)~different exchange interactions on the two crystallographically inequivalent sublattices, 
and
(3)~finite net sublattice magnetizations on the two crystallographically inequivalent sublattices.
Preconditions (1) and (2) are naturally satisfied in \NCTO\ as a result of the inequivalent crystallographic environments of the two different magnetic Co$^{2+}$ ions.
We demonstrate that Precondition (3) is naturally satisfied for the noncollinear triple-$\mathbf q$ order, but not for the collinear zigzag order.
The ferrimagnetic behavior observed in \NCTO\ thus renders the single-$\mathbf q$ zigzag scenario unlikely and points to a noncollinear triple-$\mathbf q$ ground state, consistent with the single-crystal neutron-scattering analysis~\cite{krueger23}.

The proposed mechanism is schematically depicted in Fig.~\ref{fig:schematic}.
Below the critical temperature $\Tc$, the magnetic moments on the two crystallographically inequivalent sublattices $A$ and $B$ acquire finite out-of-plane  magnetizations $m_A$ and $m_B$ with opposite signs, say $m_A > 0$ (red curve) and $m_B < 0$ (blue curve) [Precondition (3)].
Their magnitudes will generically be different, $|m_A| \neq |m_B|$ (red and dashed blue curves).
This is due to
the different $g$-factors [Precondition (1)]
and 
the different exchange interactions on the two sublattices [Precondition (2)].
Importantly, while the former effect corresponds to only simple (i.e., temperature-independent) rescalings of the sublattice magnetization curves, the latter effect generically changes the temperature dependences. This implies that the two effects can compensate each other, leading to a sign change of the net out-of-plane magnetization $m_A + m_B$ at a compensation point $\Tstar < \Tc$, as observed in \NCTO~\cite{yao20}.

The remainder of this paper is organized as follows: In Sec.~\ref{sec:model}, we describe our model. Algorithmic details of the Monte Carlo simulations and an overview of the observables considered are given in Sec.~\ref{sec:monte-carlo-simulations}. Section~\ref{sec:ferrimagnetism} contains the discussion of the finite-temperature ferrimagnetism arising from a noncollinear triple-$\mathbf{q}$ ground state, and the comparison with the collinear single-$\mathbf{q}$ zigzag case. Our conclusions are given in Sec.~\ref{sec:conclusions}.

\section{Model}
\label{sec:model}

We begin by constructing a toy model to capture the qualitative experimental features of \NCTO. Our model is composed of four key components:
\begin{enumerate}[(1), itemsep=0ex]
\item A Heisenberg-Kitaev-$\Gamma$-$\Gamma'$ term at the nearest-neighbor level.
\item A next-nearest-neighbor term that parametrizes explicit sublattice symmetry breaking.
\item A nonbilinear term that facilitates tuning between collinear single-$\mathbf{q}$ zigzag order and noncollinear triple-$\mathbf{q}$ order.
\item An onsite term that captures the effect of a weak training field.
\end{enumerate}
%
Third-neighbor interactions, which have also been proposed in the context of \NCTO~\cite{winter22}, may also be considered. However, these interactions do not alter the symmetry of the model or its ground state and, therefore, do not qualitatively affect the results. For simplicity, we neglect them from the outset.
In the following, we discuss each interaction term individually.

\subsection{Nearest neighbors: Heisenberg-Kitaev-\(\boldsymbol{\Gamma}\)-\(\boldsymbol{\Gamma'}\)}

The nearest-neighbor Heisenberg-Kitaev-$\Gamma$-$\Gamma'$ Hamiltonian consists of all bilinear spin exchange interactions compatible with the $C_3^*$ symmetry of combined spin and lattice rotations~\cite{chaloupka13, rau14a, rau14b, chaloupka15},
\begin{equation}
\label{eq:extended-KH-term}
    \begin{split}
    \mathcal H^{(1)} = &\sum_{\gamma=x,y,z} \sum_{\langle ij \rangle_{\gamma}} 
    \Bigl[ J \mathbf{S}_i \cdot \mathbf{S}_j + K S_{i}^{\gamma}S_{j}^{\gamma}\\
    & + \Gamma(S_{i}^{\alpha} S_{j}^{\beta} + S_{i}^{\beta}S_{j}^{\alpha}) \\
    & + \Gamma'(S_{i}^{\gamma}S_{j}^{\alpha} + S_{i}^{\alpha}S_{j}^{\gamma} + S_{i}^{\gamma}S_{j}^{\beta} + S_{i}^{\beta}S_{j}^{\gamma}) \Bigr]\,.
    \end{split}
\end{equation}
Here, nearest neighbors along a $\gamma$ bond on the honeycomb lattice are labeled by $\langle ij\rangle_\gamma$, with $(\alpha,\beta,\gamma)=(x,y,z)$ or related cyclic permutations.
$(S^x_i, S^y_i, S^z_i)$ are the spin components at site $i$ in the cubic coordinate system~\cite{janssen19},
with $\mathbf{S}_i = S^x_i \mathbf{e}_x + S^y_i \mathbf{e}_y + S^z_i \mathbf{e}_z$.
$J$ corresponds to the isotropic Heisenberg exchange coupling, $K$ parametrizes the Kitaev exchange, and $\Gamma$ and $\Gamma'$ characterize off-diagonal couplings.

In order to understand the magnetic long-range orders stabilized in such large parameter space, it is useful to consider isolated points that feature additional hidden symmetries. In particular, we are interested in points in parameter space in which the full SU(2) spin rotational symmetry is recovered, although not obviously so. The Heisenberg-Kitaev-$\Gamma$-$\Gamma'$ model features five classes of such hidden-SU(2)-symmetric points~\cite{chaloupka15}. For such parameter sets, the model can be mapped to a Heisenberg model in a new basis by means of a duality transformation. In fact, the recent analysis of single-crystal neutron scattering data has suggested the proximity of a minimal model for \NCTO\ to one of these special points, characterized by the couplings $(J, K, \Gamma, \Gamma')_\text{SU(2)} = (-1/9, -2/3, 8/9,-4/9) J_0$, where $J_0 > 0$ is an overall energy scale~\cite{krueger23}.
At the hidden-SU(2)-symmetric point, the ground-state manifold consists of both collinear single-$\mathbf q$ zigzag states as well as noncollinear triple-$\mathbf q$ states, which makes it easy to tune between the two experimentally scenarios by means of a suitable perturbation.
However, at this special point in parameter space, the low-temperature sublattice magnetization turns out to be exactly zero for all states within the ground-state manifold.
In order to model the generic behavior in the vicinity, but not right at, the hidden-SU(2)-symmetric point, we shift the parameters slightly. In particular, we assume an increased $\Gamma$ interaction. A positive $\Gamma>0$ acts as an antiferromagnetic coupling for spins aligned along the out-of-plane directions, but as a ferromagnetic coupling for spins aligned along the in-plane directions~\cite{janssen17, janssen19}. We find that it enhances the sublattice magnetization and concomitantly the ferrimagnetic response, this way significantly simplifying our numerics.
In our simulations, we choose the parameters
\begin{equation}
    \label{eq:hkgg'-ferrimagnetism-couplings}
    (J,K,\Gamma,\Gamma')=(-1/5,-2/3,3,-1/2) J_0,
\end{equation}
where we have absorbed a possible shift in the Kitaev coupling $K$ in a redefinition of the overall energy scale $J_0>0$.

We emphasize that the qualitative picture presented below does not depend on the particular choice of $J,K,\Gamma,\Gamma'$ within a certain range around the hidden-SU(2)-symmetric point.
%
In particular, we expect the mechanism for ferrimagnetism to also be effective for a model with a smaller $\Gamma$ interaction, such as the fitted model in Ref.~\cite{krueger23}, though the response will likely be of smaller magnitude.

\subsection{Next-nearest neighbors: Sublattice symmetry breaking}

\NCTO\ features two crystallographically inequivalent $\mathrm{Co}^{2+}$ magnetic ions, indicating that the symmetry between the two sublattices $A$ and $B$ of the honeycomb lattice is explicitly broken in the material~\cite{yao20}.
Intrasublattice interactions, such as next-nearest-neighbor terms, will therefore generically be different on the two sublattices.
For simplicity, we consider only Heisenberg next-nearest-neighbor interactions, parametrized by the couplings $J_2^A$ and $J_2^B$,
\begin{equation}
    \label{eq:sublattice-term}
    \mathcal{H}^{(2)} = J_{2}^A \sum_{\llangle ij \rrangle^A} \mathbf{S}_i \cdot \mathbf{S}_j + J_{2}^B \sum_{\llangle ij \rrangle^B} \mathbf{S}_i \cdot \mathbf{S}_j\,.
\end{equation}
Here, $\llangle ij \rrangle^A$ and $\llangle ij \rrangle^B$ correspond to next-nearest neighbors on the $A$ and $B$ sublattices, respectively.
For our simulations, we fix $J_2^{A}=-J_0/4$ and $J_2^{B}=J_0/4$. We have explicitly verified that the qualitative behavior described below is stable within a range of values of the couplings. 
%
In particular, a ferrimagnetic response will also arise if $J_2^A$ and $J_2^B$ share the same sign, as long as they differ in magnitude.
Bond-dependent interactions can emerge on the next-nearest-neighbor level as well~\cite{rousochatzakis15}, but are also not expected to change the qualitative physics, as long as they remain small.

\subsection{Nonbilinear perturbation: Six-spin ring exchange}

In the vicinity of the hidden-SU(2)-symmetric point, nonbilinear perturbations may become important~\cite{krueger23}.
Nonbilinear exchange interactions are relevant in a number of 3$d$ materials, including various chromium-, manganese-, and copper-based magnets~\cite{kvashnin20, fedorova15, dallapiazza12, larsen19}.
Here, we consider the six-spin ring exchange interaction, which arises as the leading correction to the nearest-neighbor Heisenberg exchange in the strong-coupling expansion of the single-band Hubbard model on the honeycomb lattice \cite{yang12},
\begin{equation}
\label{eq:ring-term}
    \begin{split}
       \mathcal H^{(3)} & = \frac{J_{\hexagon}}{6} \sum_{\langle ijklmn \rangle} 
        \bigl[ 2(\mathbf{S}_i \cdot \mathbf{S}_{j})(\mathbf{S}_k \cdot \mathbf{S}_{l})(\mathbf{S}_m \cdot \mathbf{S}_{n}) \\
        &\quad - 6(\mathbf{S}_i \cdot \mathbf{S}_{k})(\mathbf{S}_j \cdot \mathbf{S}_{l})(\mathbf{S}_m \cdot \mathbf{S}_{n}) \\
        &\quad + 3(\mathbf{S}_i \cdot \mathbf{S}_{l})(\mathbf{S}_j \cdot \mathbf{S}_{k})(\mathbf{S}_m \cdot \mathbf{S}_{n}) \\
        &\quad + 3(\mathbf{S}_i \cdot \mathbf{S}_{k})(\mathbf{S}_j \cdot \mathbf{S}_{m})(\mathbf{S}_l \cdot \mathbf{S}_{n}) \\
        &\quad - (\mathbf{S}_i \cdot \mathbf{S}_{l})(\mathbf{S}_j \cdot \mathbf{S}_{m})(\mathbf{S}_k \cdot \mathbf{S}_{n}) \\
        &\quad + \text{cyclic permutation of }(ijklmn) \bigr]\,,
    \end{split}
\end{equation}
where the sum runs over the elementary hexagonal plaquettes following the sites $(ijklmn)$ in counterclockwise order.
In principle, many other nonbilinear perturbations compatible with the $C_3^*$ symmetry of the Heisenberg-Kitaev-$\Gamma$-$\Gamma'$ model are admissible. However, the qualitative physical picture is expected to remain the same as long as the perturbations remain small~\cite{krueger23, wang23}.
The effect of the ring exchange perturbation can be understood in the rotated dual spin frame~\cite{francini24}.
In particular, the sign of the six-spin ring exchange interaction strongly affects the low-temperature magnetic order. By comparing the classical energies of static spin configurations, it can be shown that close to the hidden-SU(2)-symmetric point and at low temperatures, the single-$\mathbf q$ zigzag state is stabilized for $J_{\hexagon} > 0$, while the triple-$\mathbf q$ state is stabilized for $J_{\hexagon} < 0$~\cite{krueger23,francini24}.

\subsection{On site: Training field}

The three exchange interaction parts of the Hamiltonian $\mathcal H^{(1)}$, $\mathcal H^{(2)}$, and $\mathcal H^{(3)}$ are invariant under time reversal, which maps $\mathbf S_i \mapsto - \mathbf S_i$.
This implies that magnetically-ordered states come in two energetically degenerate time-reversal-related domains, both experimentally, as well as in our Monte Carlo simulations satisfying the detailed balance condition. The two domains differ in the sign of the out-of-plane magnetization component $m_A + m_B$.
In order to select one of the two domains experimentally, the sample is cooled in a weak out-of-plane training field~\cite{yao20}. At temperatures slightly below the critical temperature, the training field selects the thermodynamically-stable domain with positive out-of-plane component $m_A + m_B$. Decreasing the temperature freezes the sample in this domain even if $m_A + m_B$ changes sign below the compensation point $\Tstar$, and the frozen domain becomes only a metastable one.
In order to model this hysteretic effect in our simulations, we add a weak out-of-plane training field $\mathbf{h} = \frac{h}{\sqrt{3}}(\mathbf{e}_x + \mathbf{e}_y + \mathbf{e}_z)$ that couples to the \emph{spin moment} $\sum_i \mathbf S_i$ directly,
\begin{align}
\label{eq:external-field-term}
    \mathcal{H}^{(0)} = - \mathbf{h} \cdot \sum_i \mathbf{S}_i\,.
\end{align}
Note that the definition of the training field is different from the physical Zeeman field, which couples to the \emph{magnetic moment} $\mathbf m \propto \sum_i g_i \mathbf S_i$, where $g_i$ is the $g$-tensor at site $i$.
As a consequence, the training field selects the domain with positive out-of-plane spin component $m_A/g_A + m_B/g_B$, where $g_A$ and $g_B$ are the out-of-plane components of the $g$-tensor on the sublattices $A$ and $B$, respectively.
This allows us to fix this domain throughout the whole temperature range, even in the cases when the out-of-plane magnetization component $m_A + m_B$ is negative, i.e., for temperatures below the compensation temperature $\Tstar$.

\subsection{Full Hamiltonian}

To summarize, in this work we simulate the full Hamiltonian
\begin{align} \label{eq:full-model}
    \mathcal H = \sum_{a = 0}^3 \mathcal H^{(a)}
\end{align}
on the honeycomb lattice with $N=2L^2$ sites, spanned by the primitive lattice vectors $\mathbf a_1 = (3/2, \sqrt{3}/2)$ and $\mathbf a_2 = (3/2, -\sqrt{3}/2)$, and periodic boundary conditions.
The model includes on-site ($a=0$), nearest-neighbor ($a=1$), next-nearest-neighbor ($a=2$), and nonbilinear ring exchange ($a=3$) terms, with in total eight free parameters. We choose two different sets as follows:

\paragraph{Triple-$\mathbf q$ model.}
As a representative parameter set for a model with triple-$\mathbf q$ ground state, we use
\begin{multline}
    (J, K, \Gamma, \Gamma', J_2^A, J_2^B, J_{\hexagon} S^4) = \\
    (-1/5, -2/3, 3, -1/2, -1/4, 1/4, - 1/5)J_0, 
\end{multline}
and various values of the training-field strength $h$, allowing us to extrapolate towards the low-field limit.

\paragraph{Zigzag model.}
As a representative parameter set for a model with zigzag ground state, we use 
\begin{multline}
    (J, K, \Gamma, \Gamma', J_2^A, J_2^B, J_{\hexagon} S^4) = \\
    (-1/5, -2/3, 3, -1/2, -1/4, 1/4, 1/5)J_0, 
\end{multline}
which differs from the triple-$\mathbf q$ model only in the sign of the ring-exchange coupling $J_{\hexagon}$. This distinction allows us to explore the effects arising from the different natures of the ground states in the two models, with all bilinear couplings remaining the same. For the training field, we again simulate different field strengths $h$ in order to extrapolate towards the low-field limit.

\section{Monte Carlo simulations}
\label{sec:monte-carlo-simulations}

\subsection{Algorithmic details}

For our classical simulations, the spins are treated as three-dimensional vectors $\mathbf{S}_i=(S_i^x, S_i^y, S_i^z)$ of fixed length $S$.
We choose units in which $\mu_\mathrm{B} = \hbar = k_\mathrm{B} = J_0 = S = 1$, i.e., 
the magnetic moment $m$ is measured in units of $\mu_\mathrm{B} S$,
the temperature $T$ in units of $J_0 S^2/k_\mathrm{B}$,
the bilinear exchange couplings $J$, $K$, $\Gamma$, etc.\ in units of $J_0$,
the ring exchange coupling $J_{\hexagon}$ in units of $J_0/S^4$,
and the training field $h$ in units of $J_0 S/\mu_\mathrm{B}$.

We employ classical Monte Carlo simulations based on the Metropolis algorithm combined with a parallel tempering update with a ratio of 1:1. The typical order of magnitude of the accumulated statistics is $\mathcal O(10^5)$ configurations per point in parameter space, with each of these configurations taken after ten complete Metropolis and parallel tempering updates on the full lattice. 
Ground-state spin configurations are obtained by further cooling randomly selected Monte Carlo snapshots from a low-temperature simulation; see Ref.~\cite{janssen16} for details on the iterative minimization algorithm.
We consider $L \times L$ lattices with linear system sizes between $L = 12$ and $L=54$, increasing consecutive sizes by a constant step $\Delta L=6$. Note that the ring exchange perturbation includes 26 independent terms per plaquette, each one of it involving products of six spins. This significantly slows down the simulations in comparison with previous works on related bilinear-only models~\cite{price12, price13, chern17, janssen16, janssen17, andrade20}.

\subsection{Observables}

The following observables are measured in the simulations:

\paragraph{Static spin structure factor.}

In order to characterize the low-temperature magnetic order, we compute the static spin structure factor given by
\begin{equation}
\mathcal S({\mathbf q}) = \frac{1}{N} \sum_{i,j} \langle \mathbf{S}_i \cdot \mathbf{S}_j \rangle \, \rme^{-\rmi \mathbf q \cdot (\mathbf R_i - \mathbf R_j)},
\end{equation}
where $\mathbf{R}_{i}$ is the lattice vector at site $i$.

\paragraph{Sublattice magnetizations.}

The out-of-plane sublattice magnetizations $m_{A}$ and $m_B$ on the two sublattices $A$ and $B$ are defined as
\begin{align}
\label{eq:sublattice-magnetizations}
m_A & \coloneqq \frac{g_A}{N} \sum_{i \in A} \langle \mathbf{S}_i \cdot \mathbf{c}^* \rangle, 
&
m_B & \coloneqq \frac{g_B}{N} \sum_{i \in B} \langle \mathbf{S}_i \cdot \mathbf{c}^* \rangle,
\end{align}
where $\mathbf{c}^* = \frac{1}{\sqrt{3}} (\mathbf e_x + \mathbf e_y + \mathbf e_z)$ is the normalized out-of-plane vector and $g_A$ and $g_B$ are the out-of-plane $g$-factors on the two sublattices.
Note that the $g$-factors act only as prefactors in the above definitions. The numerical simulations are therefore independent of the specific values of $g_A$ and $g_B$, allowing these parameters to be selected a posteriori during data analysis.

Importantly, the presence of a weak out-of-plane training field $h$ in our simulations lifts the degeneracy between the two time-reversal-related domains with positive and negative, respectively, out-of-plane spin components on each of the two sublattices. 
The spin expectation values in Eq.~\eqref{eq:sublattice-magnetizations} are therefore finite already on finite lattices, in contrast to the case without external fields, in which they average to zero as a consequence of equal domain population in the unbiased simulations.

\paragraph{Sublattice Binder cumulants.}

From the sublattice magnetizations, we can construct corresponding renormalization-group-invariant Binder cumulants
\begin{equation}
    \label{eq:sublattice-binder-cumulant}
    U_{\mu}=\frac{\left\langle ( \sum_{i \in \mu} \mathbf{S}_i \cdot \mathbf{c}^* )^4 \right\rangle}{\left\langle ( \sum_{i \in \mu} \mathbf{S}_i \cdot \mathbf{c}^* )^2 \right\rangle^2}
    \quad\mathrm{with}\quad
    \mu = A,B,
\end{equation}
which are used to extract the location of the transition points from the finite-size simulations. 
If the thermal transition at $\Tc$ is continuous, the Binder cumulants are expected to scale as
\begin{equation}
    \label{eq:RG-invariant-critical-scaling}
    U_\mu=f_{R}((T-\Tc)L^{1/\nu}) + \mathcal O(L^{-\omega}),
\end{equation}
where $f_R$ is an (up to a rescaling of its argument) universal scaling function, $\nu$ is the correlation-length exponent, and $\mathcal O(L^{-\omega})$ corresponds to corrections to scaling.
If the latter can be neglected, the curves for $U_\mu$ as function of temperature $T$ for different fixed lattice sizes will cross at $\Tc$. In order to take possible corrections to scaling into account, we extrapolate the crossing points of consecutive system sizes $L$ and $L+\Delta L$ towards the thermodynamic limit $1/L \to 0$ using a power-law fit.

\paragraph{Net magnetization.}

The out-of-plane component of the net magnetization can be constructed from the sublattice magnetizations as
\begin{align}
\label{eq:total-magnetization}
M = m_A + m_B  
=  g_A \left( \frac{m_A}{g_A} \right) \left( 1+ \frac{g_B}{g_A} \frac{(m_B/g_B)}{(m_A/g_A)} \right),
\end{align}
where the ratios $m_A / g_A$ and $m_B/g_B$ denote the out-of-plane spin moments on the two sublattices, which are the actual quantities measured in the simulations. 
Typically, the sublattice magnetizations are monotonic functions of the temperature, thus the prefactor $g_A (m_A/g_A)$ acts as an overall scale. On the other hand, the quantity
\begin{equation*}
    1+ \frac{g_B}{g_A} \frac{(m_B/g_B)}{(m_A/g_A)}
\end{equation*}
is generically nonmonotonic if $m_A/g_A$ and $m_B/g_B$ have different temperature dependences.
As a consequence, while the simulations themselves are independent of the $g$-factors, the behavior of the net magnetization $M$, and thus the ferrimagnetic response of the full model, crucially depends on the ratio $g_B/g_A$.

\section{Results}
\label{sec:ferrimagnetism}

\subsection{Ground state}
\label{subsec:groundstate}

\begin{figure}[tb]
\centering
\includegraphics[width=\linewidth]{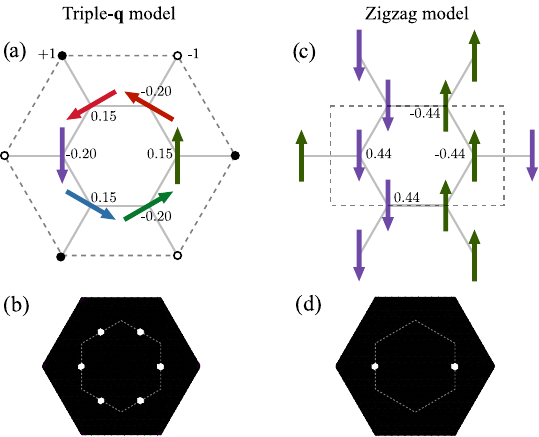}
\caption{%
(a)~Ground-state spin configuration in the triple-$\mathbf q$ model at zero field. Arrows show in-plane spin components. Filled and open dots represent spins oriented parallel and antiparallel, respectively, to the out-of-plane direction. Dashed hexagon indicates the magnetic unit cell. Numbers refer to out-of-plane spin components. The noncollinear ground state exhibits different out-of-plane magnetizations on the two crystallographic sublattices, resulting in a finite residual net magnetization.
(b)~Corresponding static spin structure factor for the triple-$\mathbf q$ configuration in (a), highlighting three Bragg peaks in the first Brillouin zone (dashed hexagon). 
(c)~Same as~(a), but for the zigzag model. The collinear ground state features a four-site unit cell (dashed rectangle) with two pairs of same-sublattice sites having opposite spin directions, resulting in zero sublattice magnetizations and vanishing total magnetization.
(d)~Same as~(b), but for the zigzag configuration in (c), revealing a single Bragg peak in the first Brillouin zone (dashed hexagon).
}
\label{fig:groundstate}
\end{figure}

\begin{figure*}[tbp]
\centering
\includegraphics[width=\textwidth]{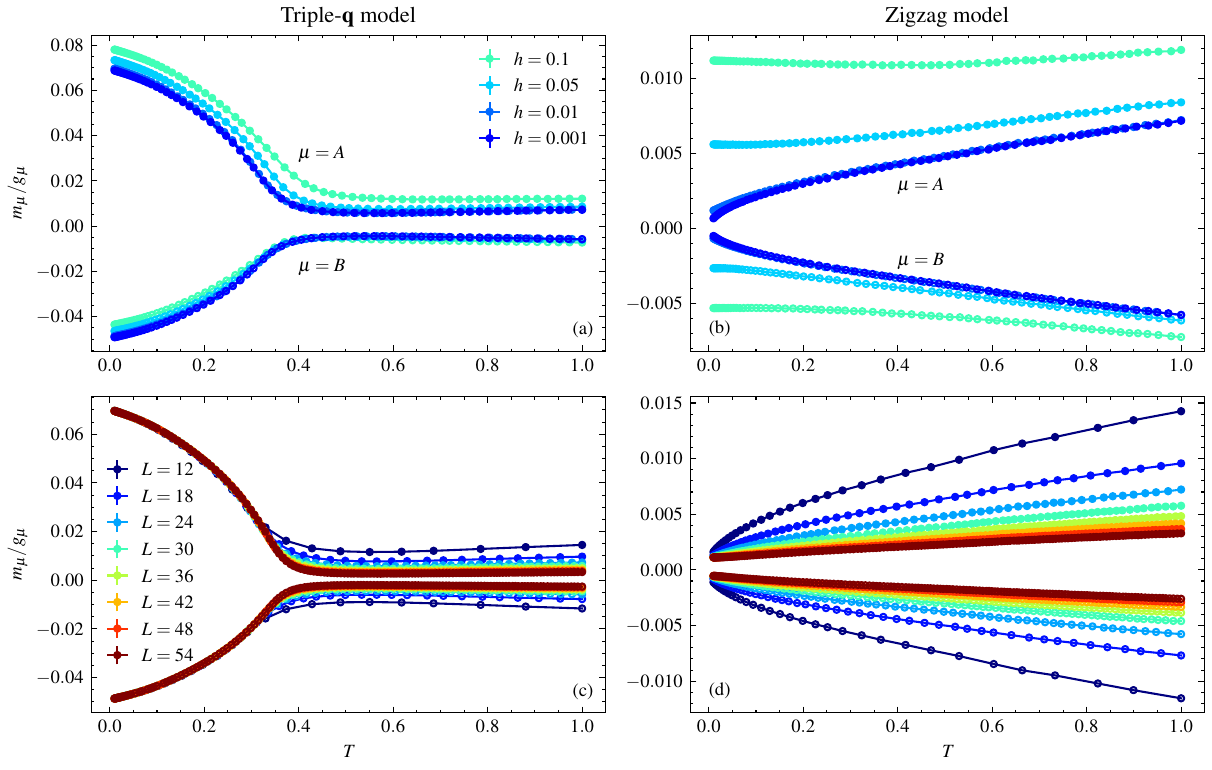}
\caption{%
(a)~Out-of-plane sublattice spin moments $m_A/g_A$ and $m_B/g_A$ in the triple-$\mathbf q$ model as function of temperature $T$ for fixed lattice size $L=24$ and various field strengths ranging from $h=0.1$ (cyan) to $h = 0.001$ (blue).
(b)~Same as~(a), but in the zigzag model. Panels (a) and (b) indicate that $h=0.01$ may be taken as a representative value for the low-field limit in both models.
(c)~Out-of-plane sublattice spin moments $m_A/g_A$ and $m_B/g_B$ in the triple-$\mathbf q$ model as function of temperature $T$ for fixed field strength $h=0.01$ and various lattice sizes ranging from $L=12$ (dark blue) to $L=54$ (dark red), indicating that the triple-$\mathbf q$ state is characterized by finite sublattice magnetizations in the thermodynamic limit.
(d)~Same as~(c), but in the zigzag model, indicating that the zigzag phase is characterized by vanishing sublattice magnetizations in the thermodynamic limit.
}
\label{fig:sublattice-magns}
\end{figure*}

Figure~\ref{fig:groundstate}(a) shows the classical ground state of the triple-$\mathbf q$ model, obtained by cooling a Monte Carlo snapshot from a simulation at temperature $T=0.01$.
The arrow lengths represent the magnitudes of the in-plane spin components. Consistent with similar models featuring negative ring exchange coupling~\cite{krueger23, wang23, francini24}, the ground state exhibits noncollinear triple-$\mathbf q$ order, as evidenced by the three Bragg peaks in the static spin structure factor shown in Fig.~\ref{fig:groundstate}(b).
Notably, the out-of-plane sublattice magnetizations on the two crystallographic sublattices differ slightly, resulting in a finite residual out-of-plane net magnetization in the triple-$\mathbf q$ state.

By contrast, the zigzag model has a collinear zigzag ground state, characterized by a single Bragg peak in the first Brillouin zone, see Figs.~\ref{fig:groundstate}(c) and \ref{fig:groundstate}(d). Its four-site magnetic unit cell consists of two pairs of same-sublattice sites with opposite spin directions, leading to zero sublattice magnetizations.
This results in a vanishing total magnetization and the absence of a ferrimagnetic response in the zigzag model, as demonstrated explicitly below.

\subsection{Sublattice spin moments}
\label{subsec:spin-moments}

In order to obtain the physics in the limit of vanishing out-of-plane training fields, the system is simulated for several decreasing field strengths.
Figure~\ref{fig:sublattice-magns}(a) shows the sublattice spin moments $m_A/g_A$ and $m_B/g_B$ as function of temperature $T$ in the triple-$\mathbf q$ model, for a fixed lattice size $L=24$ and various field strengths $h$ ranging from $0.1$ to $0.001$.
For comparison, Fig.~\ref{fig:sublattice-magns}(b) shows the same quantities in the zigzag model.
In both cases, the discrepancy between the $h=0.01$ and $h=0.001$ curves is barely noticeable. We therefore consider $h=0.01$ as representative of the low-field limit in the following analysis.

\begin{figure*}[tbp]
\centering
\includegraphics[width=\textwidth]{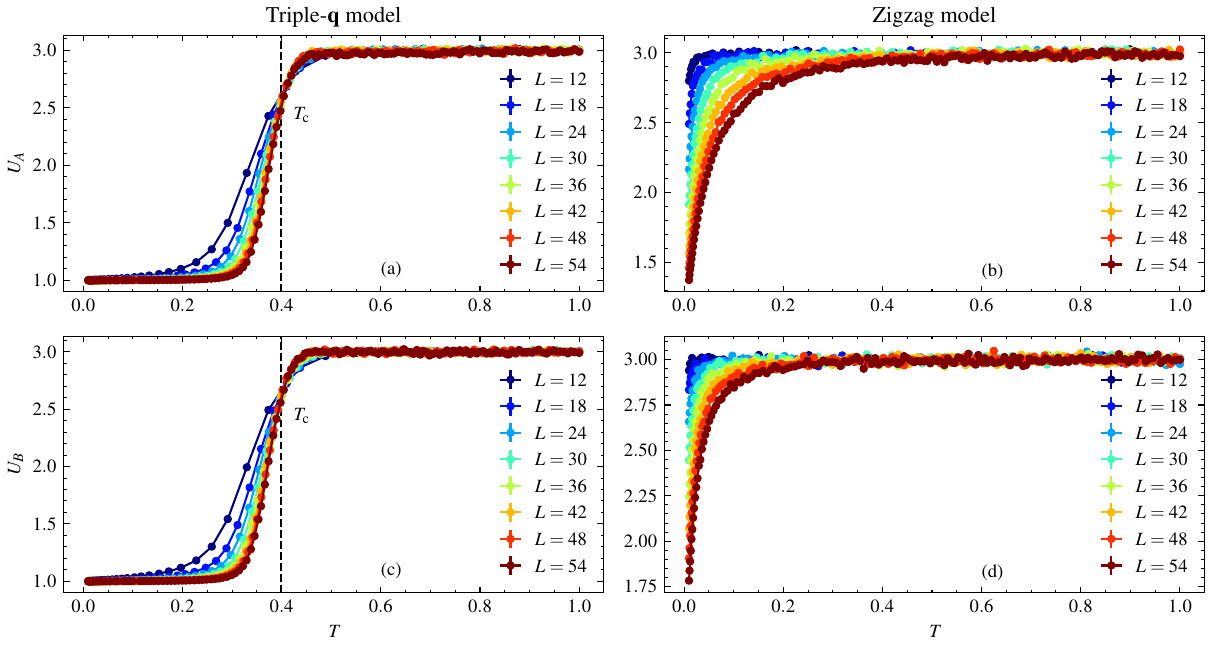}
\caption{%
(a)~Sublattice Binder cumulant $U_A$ in the triple-$\mathbf q$ model as function of temperature $T$ for various lattice sizes ranging from $L=12$ (dark blue) to $L=54$ (dark red). The crossing point indicates the critical temperature $\Tc$, marked by the vertical dashed line.
(b)~Same as~(a), but for the zigzag model. The absence of a crossing point aligns with the fact that the sublattice magnetization $m_A$ vanishes across the entire temperature range.
(c)~Same as~(a), but for the sublattice Binder cumulant $U_B$, indicating the onset of sublattice magnetization at the same critical temperature $\Tc$. 
(d)~Same as~(b), but for the sublattice Binder cumulant $U_B$, consistent with a vanishing sublattice magnetization $m_B$ across the entire temperature range.
}
\label{fig:sublattice-binder}
\end{figure*}

The system-size dependences of the sublattice spin moment curves is illustrated in Fig.~\ref{fig:sublattice-magns}(c) for the triple-$\mathbf q$ model and in Fig.~\ref{fig:sublattice-magns}(d) for the zigzag model. 
In both cases, the spin moment $m_A/g_A$ on the $A$ sublattice is slightly larger in magnitude than that on the $B$ sublattice across the entire temperature range. As a result, the out-of-plane training field $\mathbf{h}$ leads to the lifting of degeneracy between time-reversal-related domains in a way that $m_A > 0$ and $m_B < 0$.
Furthermore, in both cases, the magnitudes of the sublattice spin moments decrease with increasing system size at sufficiently high temperatures, suggesting they vanish for infinitesimally small $h$ in the thermodynamic limit, consistent with the expectation in a paramagnetic phase above a finite $\Tc$.
At low temperatures, however, the behavior of the sublattice spin moments differs significantly between the two models.
In the triple-$\mathbf q$ model [Fig.~\ref{fig:sublattice-magns}(c)], below a finite $\Tc \simeq 0.4$, the spin moment curves extrapolate to finite values in the thermodynamic limit, indicating that the triple-$\mathbf q$ ground state of this model is characterized by finite sublattice magnetizations.
In the zigzag model [Fig.~\ref{fig:sublattice-magns}(d)], by contrast, the magnitudes of the sublattice spin moments continue to decrease with increasing system size throughout the whole temperature range, indicating that the zigzag ground state of this model is characterized by vanishing sublattice magnetizations, in agreement with the discussion in Sec.~\ref{subsec:groundstate}.
This result aligns with the behavior observed in other extended Heisenberg-Kitaev models with a zigzag ground state: in all known examples, such states are collinear, precluding their coexistence with N\'eel order and resulting in the absence of any ferrimagnetic response.

\subsection{Critical temperature}

The behavior of the sublattice spin moments in the triple-$\mathbf q$ model suggest a phase transition at around $\Tc \simeq 0.4$. In order to determine the critical temperature in the thermodynamic limit, we study the finite-size scaling of the sublattice Binder cumulants $U_\mu$ with $\mu = A,B$.

Figures~\ref{fig:sublattice-binder}(a) and \ref{fig:sublattice-binder}(c) show the sublattice Binder cumulants $U_A$ and $U_B$, respectively, in the triple-$\mathbf q$ model as function of temperature $T$ for various fixed lattice sizes ranging from $L=12$ to $L=54$. The crossing points of the curves for consecutive system sizes depend only weakly on system size. Extrapolating the crossing points towards $1/L \to 0$ using a polynomial fit up to third order leads to the critical temperature
\begin{equation}
    \label{eq:critical-temperature}
    \Tc = 0.399(5)
\end{equation}
in the thermodynamic limit, for both $U_A$ and $U_B$.
This temperature corresponds to the N\'eel temperature at which the system enters an out-of-plane ordered phase. 
Alternatively, the critical temperature can be obtained through Bayesian-inference-based finite-size scaling collapses of Binder cumulants~\cite{harada11, francini24}. We have explicitly verified that this approach leads to a result for the critical temperature that is consistent with Eq.~\eqref{eq:critical-temperature}.

For comparison, the sublattice Binder cumulants in the zigzag model are shown in Figs.~\ref{fig:sublattice-binder}(b) and \ref{fig:sublattice-binder}(d). The absence of a crossing point in the curves for both sublattices aligns with the fact that both sublattice magnetizations vanish in the zigzag state across the entire temperature range in the thermodynamic limit, in agreement with what has been argued in Secs.~\ref{subsec:groundstate} and \ref{subsec:spin-moments}.
Note that the absence of a crossing point in the sublattice Binder cumulants does not indicate the absence of a well-defined ordering temperature in the zigzag model. This temperature can be determined from the crossings of Binder cumulants associated with the zigzag staggered magnetization (not shown).

\subsection{\texorpdfstring{$\boldsymbol{g}$}{g}-factors}

Since the sublattice magnetizations vanish across the entire temperature range in the zigzag model, leading to the absence of ferrimagnetic behavior, we will, in what follows, focus exclusively on the triple-$\mathbf q$ model, which features a low-temperature order with finite sublattice magnetizations.
In the triple-$\mathbf q$ model, the net spin moment $m_A/g_A+m_B/g_B$ is nonnegative and monotonically increases with decreasing temperature (not shown). This, however, does not imply that the experimentally readily observable residual magnetization $M = m_A + m_B$ is monotonic as function of temperature as well.
This is because due to the two inequivalent crystallographic sublattices $A$ and $B$ in \NCTO, the corresponding $g$-factors $g_A$ and $g_B$ can differ.
For $m_A > 0$ and $m_B < 0$, as in Fig.~\ref{fig:sublattice-magns}, the magnetization $M$ in units of $g_A$ can be written as
\begin{equation}
    \label{eq:magnetization-ratio}
    \frac{M}{g_A}= \frac{m_A}{g_A} \left( 1 - \frac{g_B}{g_A} \frac{|m_B/g_B|}{|m_A/g_A|} \right),
\end{equation}
cf.\ Eq.~\eqref{eq:total-magnetization}.
Since $g_A$ and $g_B$ are positive, the sign of $M$ depends on whether the ratio of sublattice moments $|m_A/g_A| / |m_B/g_B|$ is larger or smaller than the $g$-factor ratio $g_B/g_A$. In fact, from Eq.~\eqref{eq:magnetization-ratio}, we find that
\begin{equation}
\sgn M = 
    \begin{cases}
        +1 & \text{for $|m_A/g_A|/|m_B/g_B|>g_B/g_A$}, \\
        -1 & \text{for $|m_A/g_A|/|m_B/g_B|<g_B/g_A$}.
    \end{cases}
\end{equation}
\begin{figure}[tb]
\centering
\includegraphics[width=\linewidth]{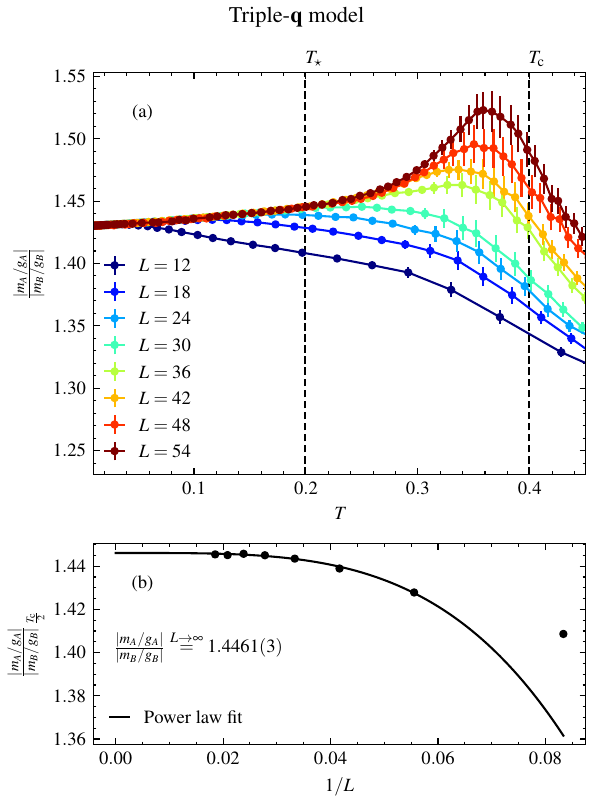}
\caption{%
(a)~Ratio of sublattice spin moments $|m_A/g_A|/|m_B/g_B|$ as function of temperature in the triple-$\mathbf q$ model for various lattice sizes ranging from $L=12$ (dark blue) to $L=54$ (dark red). For sufficiently large system sizes and temperatures not too close to $\Tc$, the ratio $|m_A/g_A|/|m_B/g_B|$ monotonically increases with temperature $T$. Note that the ratio is not well defined for $T \geq \Tc$, since both $m_A$ and $m_B$ vanish in the paramagnetic phase in the thermodynamic limit. The dashed lines indicate the critical temperature $\Tc$ and the compensation point $\Tstar = \Tc/2$.
(b)~Ratio of sublattice spin moments $|m_A/g_A|/|m_B/g_B|$ at fixed temperature $T = \Tstar = \Tc/2$ as function of $1/L$. Solid black line indicates the power-law fit according to Eq.~\eqref{eq:power-law-fit}.}
\label{fig:ratio}
\end{figure}
Importantly, the ratio of sublattice spin moments $|m_A/g_A|/|m_B/g_B|$ generically depends on temperature, and might thus cross the temperature-independent $g$-factor ratio at a particular temperature $\Tstar$. In such case, the temperature $\Tstar$ corresponds to the compensation point, at which the residual magnetization $M$ changes sign.
In Fig.~\ref{fig:ratio}(a), we show the ratio of sublattice spin moments $|m_A/g_A|/|m_B/g_B|$ as function of temperature in the triple-$\mathbf q$ model.
Note that the ratio $|m_A/g_A|/|m_B/g_B|$ is not well defined for $T \geq \Tc$, since both $m_A$ and $m_B$ vanish in the paramagnetic phase in the thermodynamic limit.
This explains the increased finite-size effects that occur when approaching the critical temperature $\Tc$ from below.
Importantly, for sufficiently large system sizes, the ratio $|m_A/g_A|/|m_B/g_B|$ monotonically increases with temperature $T$ between $|m_A/g_A|/|m_B/g_B| = 1.4302(2)$ for $T \to 0$ and $|m_A/g_A|/|m_B/g_B| = 1.522(15)$ for $T \simeq 0.36$.
As a result, a compensation point at a finite temperature $\Tstar$ below the critical temperature $\Tc$ exists in the triple-$\mathbf q$ model if, and only if, the ratio $g_B/g_A$ of the $g$-factors on the two sublattices falls within this range.
Conversely, the experimental observation of a compensation point in \NCTO\ at a temperature $\Tstar \simeq \Tc/2$~\cite{yao20} may in principle allow one to \emph{predict} the $g$-factor ratio $g_B/g_A$.
To this end, we extrapolate the ratio of sublattice moments $|m_A/g_A|/|m_B/g_B|$ at $T = \Tc/2$ towards the thermodynamic limit using a power-law ansatz as function of $1/L$,
\begin{equation}
    \label{eq:power-law-fit}
    \left.\frac{\left|m_A/g_A\right|}{\left|m_B/g_B\right|}\,\right|_{T=\Tc/2} = a - \frac{b}{L^c},
\end{equation}
with positive fitting parameters $a$, $b$, and $c$. Figure~\ref{fig:ratio}(b) presents our best fit, derived by excluding the smallest system size $L=12$, resulting in $\lim_{L \to \infty} \left|{m_A}/{g_A}\right| / \left|{m_B}/{g_B}\right| = 1.4461(3)$.
This implies that the experimentally observed compensation point at $\Tstar \simeq \Tc/2$ in \NCTO\ can be explained within the triple-$\mathbf q$ model, provided the $g$-factor on the sublattice with antiferromagnetic intrasublattice exchange ($B$ sublattice in our case, with $J_2^B>0$) in this toy model is approximately 40\% larger than the $g$-factor on the sublattice with ferromagnetic intrasublattice exchange ($A$ sublattice, with  $J_2^A<0$).
A larger ratio of $g_B/g_A$ will shift the compensation point to higher temperatures, while a smaller ratio will shift it to lower temperatures.

This result can be understood as follows: At temperatures below, but near the critical temperature $\Tc$, enhanced fluctuations on the $A$ sublattice, which has ferromagnetic intrasublattice exchange, result in a greater magnitude of sublattice magnetization compared to that on the $B$ sublattice with antiferromagnetic intrasublattice exchange. However, at lower temperatures, this effect can be offset by the increased static magnetic moment on the $B$ sublattice, due to its larger $g$-factor.

In conclusion, we anticipate that for \NCTO, the $g$-factor on the crystallographic sublattice with stronger antiferromagnetic intrasublattice exchange is larger than that on the other sublattice.
To obtain a quantitative prediction of the $g$-factor sublattice anisotropy, a more refined and rigorously validated effective spin model for \NCTO\ is required. To the best of our knowledge, such a model is currently unavailable.
Nevertheless, to proceed, we adhere to our toy model and set the $g$-factor ratio at $g_B/g_A = 1.446$. As we will demonstrate next, this choice results in a residual magnetization with the same qualitative temperature dependence as the experimental data in \NCTO.

\subsection{Residual magnetization}

\begin{figure}[tb]
\centering
\includegraphics[width=\linewidth]{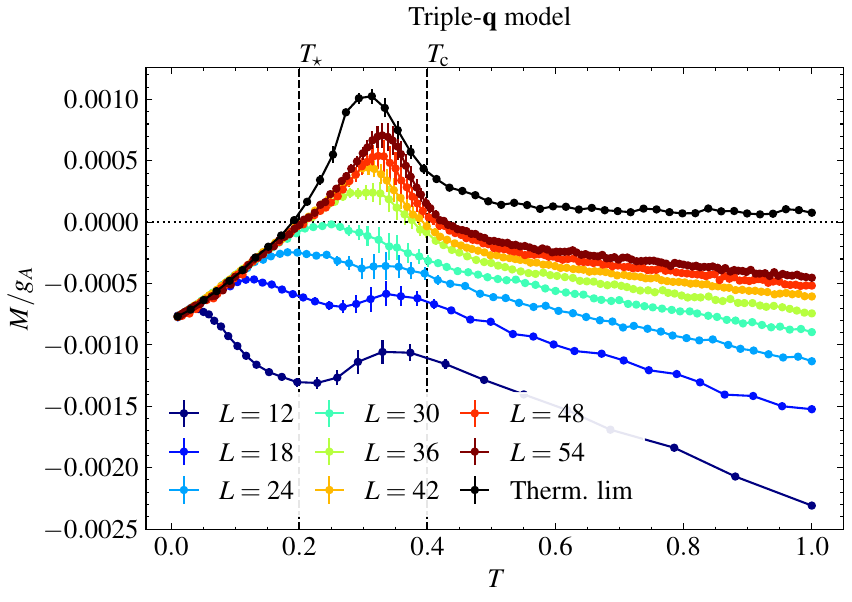}
\caption{%
Out-of-plane net magnetization $M$ in units of $g_A$ as function of temperature $T$ in the triple-$\mathbf q$ model for various lattice sizes ranging from $L=12$ (dark blue) to $L=54$ (dark red), assuming a $g$-factor sublattice anisotropy ratio of $g_B/g_A = 1.446$. Black dots represent the extrapolation of the finite-size data to the thermodynamic limit.
The data reveal three distinct temperature regimes: In the paramagnetic phase ($T > \Tc$), the magnetization vanishes in the thermodynamic limit. Below the transition, the out-of-plane magnetization is initially positive for $\Tstar < T < \Tc$, but changes sign at $\Tstar$ and becomes negative for $T < \Tstar$.
%
Note that finite-size effects become significant near the critical temperature $\Tc$, as expected for a continuous transition. As a result, the expected kink in the extrapolated $L \to \infty$ magnetization curve is smoothed out.
}
\label{fig:ferrimagnetism}
\end{figure}

Figure~\ref{fig:ferrimagnetism} shows the out-of-plane net magnetization $M$ in units of $g_A$ as a function of temperature $T$ for various fixed $L$. 
As the lattice size increases, the magnetization curve increasingly aligns with the experimental observations in \NCTO~\cite{yao20}, compare also Fig.~\ref{fig:schematic}:
In the low-temperature limit, a small, but finite, residual magnetization with a \emph{negative} component along the training direction remains. Upon increasing the temperature, the magnetization component gradually shifts towards zero and eventually changes sign at a compensation point $\Tstar = \Tc/2$. Above the compensation point, the magnetization component first increases to a maximum value approximately equal in magnitude but opposite in sign to that in the low-temperature limit, reaching this maximum only slightly below $\Tc$, and then drops to zero as $T$ approaches $\Tc$ from below.
In the high-temperature regime for $T > \Tc$, the magnitude of the magnetization decreases with increasing system size, suggesting that it vanishes in the thermodynamic limit.

The behavior of the magnetization $M$ in the thermodynamic limit can be obtained by using power-law fits analogous to those of Eq.~\eqref{eq:power-law-fit} for various fixed temperatures. We employ an equally-spaced grid of $50$ different temperatures between $T=0$ and $T=1$.
The resulting magnetization curve, representative for the thermodynamic limit, is shown as black line in Fig.~\ref{fig:ferrimagnetism}.
At low temperatures up to about the compensation point $\Tstar$, finite-size effects are small, and the magnetization curves on the largest lattices are representative for the thermodynamic limit.
Finite-size effects become significant as the temperature approaches the critical temperature $\Tc$, which is expected for a continuous phase transition. In particular, near $\Tc$, the magnetization curve resulting from the $L \to \infty$ extrapolation now is sensitive to the choice of fitting model. As a result, the expected kink in $M(T)$ at $T = \Tc$ is smoothed out. Nonetheless, in agreement with the experimental findings~\cite{yao20}, we observe a distinct maximum in $M$ between $\Tstar$ and $\Tc$, with $M$ dropping towards zero as $T$ approaches $\Tc$ from below.
In the high-temperature regime for $T > \Tc$, the fit indicates a finite-size dependence that is consistent with $1/L$ scaling, as expected in the paramagnetic phase.

We conclude that the ferrimagnetic response of \NCTO\ is qualitatively well described by the triple-$\mathbf q$ model.

\section{Conclusions}
\label{sec:conclusions}

In this work, we have explored the conditions that give rise to ferrimagnetic behavior, as observed in \NCTO~\cite{yao20}, within extended Heisenberg-Kitaev models.
In particular, we have explored two exemplary models within the theory space spanned by nearest-neighbor Heisenberg, Kitaev, and off-diagonal $\Gamma$ and $\Gamma'$ interactions, next-nearest-neighbor Heisenberg interactions, and non-bilinear ring interactions across hexagonal plaquettes on the honeycomb lattice.
The two different models, which differ only in the sign of the ring exchange coupling, stabilize two different ground states:
The first model, dubbed the ``zigzag model,'' features three degenerate pairs of collinear zigzag ground states related by $C_3^*$ spin-lattice rotational symmetry, with time-reversal-related partners within each pair. Each zigzag state is characterized by a single Bragg peak at one of the three $\mathbf M$ points in the first Brillouin zone.
The second model, dubbed the ``triple-$\mathbf q$ model,'' features four degenerate pairs of noncollinear triple-$\mathbf q$ states related by lattice translational symmetry, again with time-reversal-related partners within each pair. Each triple-$\mathbf q$ state is characterized by three simultaneous Bragg peaks at the three $\mathbf M$ points in the first Brillouin zone.
In real space, the vortex-like spin structure in the triple-$\mathbf q$ state, see Fig.~\ref{fig:groundstate}(a), induces a finite magnetic toroidal moment along the out-of-plane direction~\cite{ederer07}. Once the out-of-plane magnetization direction is selected by the training field, the toroidal moment's direction becomes fixed as well. Consequently, techniques sensitive to toroidal moments may detect a time-reversal-odd response below $\Tc$ without the appearance of a compensation point.

Importantly, in the zigzag state, the sublattice magnetization vanishes in the thermodynamic limit, leading to a complete absence of residual magnetization across the entire temperature range. Consequently, a model with a collinear zigzag ground state is insufficient to capture the ferrimagnetic response observed in \NCTO~\cite{yao20}.
While we have demonstrated this explicitly for the current toy model, we understand this result to be generally applicable. This is because collinear zigzag order alone, without additional N\'eel order, does not produce a net moment on either sublattice, even when the sublattice symmetry is explicitly broken.
In contrast, the noncollinear triple-$\mathbf{q}$ state naturally exhibits a finite sublattice moment when the sublattice symmetry is explicitly broken, leading to a finite residual magnetization in the low-temperature phase of our triple-$\mathbf{q}$ model.

The residual out-of-plane magnetization in \NCTO\ reverses sign as function of temperature at a compensation point approximately half the ordering temperature~\cite{yao20}.
Our triple-$\mathbf{q}$ model can account for this behavior, if the out-of-plane $g$-factor on the sublattice with stronger antiferromagnetic intrasublattice exchange is larger than that on the other sublattice. The exact ratio of $g$-factors between the two sublattices is expected to depend sensitively on the difference in intrasublattice exchange, such as the ratio of next-nearest-neighbor Heisenberg couplings.

The behavior of \NCTO\ at intermediate and high temperatures is well captured by the triple-$\mathbf q$ model. However, at very low temperatures, a qualitative discrepancy arises between the classical simulation (black curve in Fig.~\ref{fig:ferrimagnetism}) and the experiments reported in Fig.~2(d) of Ref.~\cite{yao20} (see also the schematic in Fig.~\ref{fig:schematic}).
Namely, in the low-temperature limit, the experimental magnetization curve $M(T)$ becomes essentially flat.
In \NCTO, this flattening of $M(T)$ begins around 10 K, corresponding to the inflection point of $M(T)$. This behavior is absent in our classical simulations, where the slope of $M(T)$ remains finite down to the lowest temperatures.
We attribute the flattening of $M(T)$ at low temperatures to the absence of a continuous spin rotational symmetry in \NCTO, which introduces a finite energy gap for magnetic excitations. When the temperature is much lower than the gap scale, the number of excitations is exponentially suppressed, leading to reduced temperature dependence of the magnetization.
Indeed, the gap for magnetic excitations, previously measured in neutron scattering experiments, is approximately 1 meV~\cite{yao22, krueger23}, which is consistent with the temperature scale at which the flattening of $M(T)$ begins in \NCTO.
Such effects are inherently beyond the reach of classical simulations.
As a complementary approach, the low-temperature behavior of $M(T)$ could be studied using spin-wave theory~\cite{consoli21}, which is left for future investigation.

Looking ahead, establishing a quantitatively reliable effective spin model for \NCTO\ is crucial. Our work imposes two significant constraints on such a model:
First, and maybe most importantly, it must feature a ground state that allows a finite sublattice magnetization. As demonstrated in this work, such constraint is difficult to realize, at least classically, in a model with collinear zigzag ground state, but naturally achieved with a model realizing a triple-$\mathbf q$ ground state.
Second, it must accurately account for the explicit sublattice symmetry breaking present in \NCTO. This involves balancing sublattice anisotropies related to static magnetic moments, influenced by different $g$-factors, with those related to fluctuating spin moments, determined by varying intrasublattice exchange interactions.
Achieving this balance is crucial for explaining the sign change in residual magnetization at the compensation point, which occurs in \NCTO\ at approximately half the ordering temperature.
Finding such a model would not only deepen our understanding of the magnetic behavior of \NCTO\ but also clarify the debate about the significance of bond-dependent interactions in this material~\cite{winter22, songvilay20, kim22, sanders22, krueger23, miao24}, and ideally shed light on the enigmatic behavior observed under finite magnetic fields~\cite{lin21, hong21, yang22, xiang23, hong23, pilch23, bera23, gillig23}.
 
\begin{acknowledgments}

We thank Pedro M.\ C\^onsoli and Yuan Li for valuable discussions and insightful feedback on the manuscript.
This work has been supported by the Deutsche Forschungsgemeinschaft (DFG) through SFB 1143 (A07, Project No.\ 247310070), the W\"urzburg-Dresden Cluster of Excellence \textit{ct.qmat} (EXC 2147, Project No.\ 390858490), and the Emmy Noether program (JA2306/4-1, Project No.\ 411750675). 
The authors gratefully acknowledge the computing time made available to them on the high-performance computer at the NHR Center of TU Dresden. This center is jointly supported by the German Federal Ministry of Education and Research and the state governments participating in the NHR~\cite{nhr-alliance}.

\end{acknowledgments}


The data that support the findings of this article are openly available~\cite{data-availability}.

\bibliographystyle{longapsrev4-2}
\bibliography{HK-ferri}

\end{document}